\documentclass[twocolumn,showpacs,superscriptaddress,amsmath,amssymb,aps,pra]{revtex4-2}
\usepackage{graphicx}
\usepackage{CJKutf8}
\usepackage{dcolumn}
\usepackage{amsmath,bm}
\usepackage{epstopdf}
\usepackage[mathlines]{lineno}
\usepackage{color}
\usepackage[colorlinks=true,linkcolor=blue,citecolor=magenta,urlcolor=cyan]{hyperref}

\usepackage{tikz,xcolor,hyperref}

\definecolor{lime}{HTML}{A6CE39}
\DeclareRobustCommand{\orcidicon}{%
	\begin{tikzpicture}
	\draw[lime, fill=lime] (0,0) 
	circle [radius=0.16] 
	node[white] {{\fontfamily{qag}\selectfont \tiny ID}};
	\draw[white, fill=white] (-0.068,0.105) 
	circle [radius=0.007];
	\end{tikzpicture}
	\hspace{-2mm}
}

\foreach \x in {A, ..., Z}{%
	\expandafter\xdef\csname orcid\x\endcsname{\noexpand\href{https://orcid.org/\csname orcidauthor\x\endcsname}{\noexpand\orcidicon}}
}

\begin{document}
\title{Electrical manipulation of a hole `spin'-orbit qubit in nanowire quantum dot: The nontrivial magnetic field effects}
\author{Rui\! Li~(\begin{CJK}{UTF8}{gbsn}李睿\end{CJK})}
\email{ruili@ysu.edu.cn}
\affiliation{Key Laboratory for Microstructural Material Physics of Hebei Province, School of Science, Yanshan University, Qinhuangdao 066004, China}

\author{Hang\! Zhang~(\begin{CJK}{UTF8}{gbsn}张航\end{CJK})}
\affiliation{Key Laboratory for Microstructural Material Physics of Hebei Province, School of Science, Yanshan University, Qinhuangdao 066004, China}

\begin{abstract}

Strong `spin'-orbit coupled one-dimensional hole gas is achievable in a Ge nanowire in the presence of a strong magnetic field. The strong magnetic field lifts the two-fold degeneracy in the hole subband dispersions, so that the effective low-energy subband dispersion exhibits strong `spin'-orbit coupling. Here, we study the electrical `spin' manipulation in a Ge nanowire quantum dot for both the lowest and second lowest hole subband dispersions. Using a finite square well to model the quantum dot confining potential, we calculate exactly the level splitting of the `spin'-orbit qubit and the Rabi frequency in the electric-dipole `spin' resonance. The `spin'-orbit coupling modulated longitudinal $g$-factor $g_{\rm so}$ is not only non-vanishing but also magnetic field dependent. Moreover, the `spin'-orbit couplings of the lowest and second lowest subband dispersions have opposite magnetic dependences, so that the results for these two subband dispersions are totally different.
\end{abstract}
\date{July 26, 2022}
\maketitle

\section{\label{Sec_I}Introduction}

Due to the potential applications in both quantum information technologies and spintronics~\cite{RevModPhys.76.323,RevModPhys.79.1217,WU201061,vandersypen2019,10.1093/nsr/nwy153}, hole spins confined in semiconductor quantum dots have attracted considerable interest recently~\cite{Scappucci:2021vk,Hendrickx:2020ab,Jirovec:2021uv}. The hole spin in the valence band has several additional merits in comparison with the electron spin in the conduction band~\cite{loss1998quantum,witzel2006quantum,yao2006theory,cywinski2009electron}. First, hole in the valence band has p-type Bloch function, such that the hyperfine interaction between the hole spin and the lattice nuclear spins is small~\cite{winkler2003spin}. For semiconductor Ge, this hyperfine interaction can be further suppressed through isotopic purification~\cite{Balasubramanian:2009ub}. Second, the low-energy hole states in the bulk valence band are well described by the $4\times4$ Luttinger-Kohn Hamiltonian~\cite{PhysRev.97.869,PhysRev.102.1030}, where there is a large intrinsic spin-orbit coupling. This spin-orbit coupling can facilitate fast manipulation of the hole spin in an external oscillating electric field~\cite{Wang:2022tm,Wang:2021wc,PhysRevB.103.125201,PhysRevB.103.085309,PhysRevB.105.075313}.

Before trapping a hole in a gated semiconductor quantum dot, one should first achieve two-dimensional (2D) hole gas in a quantum well~\cite{Hendrickx:2020ab,Hendrickx:2020aa} or one-dimensional (1D) hole gas~\cite{Lu10046,PhysRevLett.101.186802,Watzinger:2018aa,PhysRevResearch.3.013081} in a quantum wire. Note that quantum wire also has extensive application in the study of the spin wave physics~\cite{doi:10.1063/1.4711039,Wang:2021ts}. The large intrinsic hole spin-orbit coupling is expected to give rise to sizable effects when strong quantum confinement is present. It is well-known for 2D hole gas, due to the strong 1D confinement of the quantum well, there is a heavy-hole-light-hole anticrossing in the subband dispersions~\cite{winkler2003spin}. Nevertheless, the lowest subband dispersion is nearly parabolic, and the band minimum is at the center of $k$ space~\cite{winkler2003spin,PhysRevB.36.5887}. While for 1D hole gas, due to the strong 2D confinement of the quantum wire, the results are totally different~\cite{PhysRevB.84.195314,PhysRevB.97.235422,RL_2021_Low}. The lowest two subband dispersions can be regarded as two mutually displaced parabolic curves with an anticrossing at the center of the $k_{z}$ space~\cite{PhysRevB.84.195314,PhysRevB.97.235422,RL_2021_Low}. We note that, for both 1D and 2D hole gases, each dispersion curve is two-fold degenerate.

When we use a strong magnetic field to lift the two-fold degeneracy in the 1D hole subband dispersions, there is a strong `spin'-orbit coupling in the remaining effective low-energy subband dispersion~\cite{Li_2022_Searching}. Interestingly, the resulting low-energy subband dispersion can be modeled as $E_{0}(k_{z})=\hbar^{2}k^{2}_{z}/(2m^{*}_{h})+\alpha\sigma^{z}k_{z}+g^{*}_{h}\mu_{B}B\sigma^{x}/2$, where $\sigma^{z,x}$ are the Pauli `spin' matrices~\cite{Li_2022_Searching}. Note that this strong spin-orbit coupled 1D dispersion has many applications in the studies of the spin-orbit qubit~\cite{trif2008spin,lirui2013controlling,PhysRevB.87.205436,PhysRevB.92.054422,PhysRevApplied.14.014090,dong2019theoretical} and the Bose gases~\cite{ting2020some,zhi2019one,Feng_2020,Hai_2020,Wang_2020}. Inspired by recent advances in manipulating electrically the hole spin in a Ge nanowire quantum dot~\cite{Froning:2021aa,PhysRevResearch.3.013081,PhysRevB.99.115317,PhysRevB.104.235304,PhysRevB.105.075308}, here we address the `spin'-orbit coupling physics in the Ge nanowire.

In this paper, we study the `spin'-orbit qubit in a Ge nanowire quantum dot for both the lowest and second lowest hole subband dispersions. The nanowire quantum dot can be achieved by applying local trap potential to a 1D strong `spin'-orbit coupled hole gas. We use a finite square well to model the quantum dot confining potential. Not only is finite potential more realistic than infinite potential from the experimental point of view, but also the square well model is exactly solvable even in the presence of the strong `spin'-orbit coupling~\cite{lirui2018the,lirui2018a,Li2020charge}. Both the level splitting and the basis states of the `spin'-orbit qubit are obtained exactly. We then study the electric-dipole `spin' resonance~\cite{Nowack1430,nadj2010spin} when an oscillating electric field is present. Both the qubit level splitting and the Rabi frequency of the driving electric field are shown to have non-trivial and non-linear responses to the applied magnetic field strength. Also, the magnetic field dependence of the `spin'-orbit coupling of the lowest subband dispersion is opposite to that of the second lowest subband dispersion, such that the results for these two subband dispersions are totally different.

\section{\label{Sec_II}Nanowire hole quantum dot}

Strong `spin'-orbit coupled 1D hole gas is achievable in a cylindrical Ge nanowire in the presence of a strong external magnetic field~\cite{Li_2022_Searching}. Here, let us briefly explain the realization of this strong `spin'-orbit coupled 1D hole gas. The subband quantization of the hole gas in a cylindrical Ge nanowire is described by the following Hamiltonian~\cite{PhysRevB.79.155323,PhysRevB.84.195314}
\begin{eqnarray}
H_{0}&=&\frac{1}{2m_{e}}\left[\left(\gamma_{1}+\frac{5}{2}\gamma_{s}\right)\textbf{p}'^{2}-2\gamma_{s}(\textbf{p}'\cdot\textbf{J})^{2}\right]\nonumber\\
&&+2\kappa\mu_{B}{\bf B}\cdot{\bf J}+V(r),\label{eq_model}
\end{eqnarray}
where $m_{e}$ is the bare electron mass, $\gamma_{1}=13.35$, $\gamma_{s}=5.11$, and $\kappa=3.41$ are Luttinger parameters~\cite{PhysRevB.4.3460} in the spherical approximation, ${\bf p'}=-i\hbar\nabla+e{\bf A}$ is the momentum operator in the presence of the magnetic field, with ${\bf A}$ being the vector potential ${\bf B}=\nabla\times{\bf A}$, ${\bf J}=(J_{x}, J_{y}, J_{z})$ is a spin-$3/2$ vector operator, and $V(r)$ is the transverse ($xy$ plane) confining potential
\begin{equation}
V(r)=\left\{\begin{array}{cc}0,~&~r<R,\\
\infty,~&~r>R,\end{array}\right.\label{Eq_potential}
\end{equation}
with $R$ being the radius of the Ge nanowire. Note that in our following calculations we have set $R=10$ nm, a typical and experimentally achievable radius~\cite{PhysRevLett.101.186802,PhysRevLett.112.216806}.

Via choosing proper gauge for the vector potential, e.g., we can let ${\bf A}=(0,0,By)$ when the magnetic field ${\bf B}=(B,0,0)$ is transverse, we have the operator $p_{z}=\hbar{}k_{z}$ in Hamiltonian (\ref{eq_model}) conserved. It follows that, even in the presence of the magnetic field, Hamiltonian~(\ref{eq_model}) will give rise to a 1D dispersions $E_{n}(k_{z})$, where $n$ is the subband index. We next determine $E_{n}(k_{z})$ by treating the magnetic field as a perturbation. First, in the absence of the magnetic field ${\bf B}=0$, the Hamiltonian (\ref{eq_model}) is exactly solvable~\cite{sweeny1988hole,PhysRevB.42.3690,RL_2021_Low}. Interestingly, the lowest two subband dispersions are just two mutually displaced parabolic curves with an anticrossing at $k_{z}R=0$, and each dispersion curve is two-fold degenerate~\cite{PhysRevB.84.195314,RL_2021_Low}. Second, the strong magnetic field lifts the two-fold degeneracy in the subband dispersions.

Typical results of the subband dispersions with spin splitting are shown in Figs.~\ref{fig_fitting1} (longitudinal field) and \ref{fig_fitting2} (transverse field) in appendix~\ref{appendix_a}. Now, the four subband dispersions can be merged into two sets of combined dispersions (we label them respectively the lowest and second lowest subband dispersions), each of which can be modeled by~\cite{Li_2022_Searching}
\begin{equation}
E_{0}(k_{z})=\frac{\hbar^{2}k^{2}_{z}}{2m^{*}_{h}}+\alpha\sigma^{z}k_{z}+\frac{g^{*}_{h}\mu_{B}B}{2}\sigma^{x}+const.,\label{eq_dispersion}
\end{equation}
where $\sigma^{z,x}$ are the Pauli `spin' matrices. We also emphasize that the effective hole mass $m^{*}_{h}\equiv{}m^{*}_{h}(B)$, the Rashba type `spin'-orbit coupling~\cite{bychkov1984oscillatory} $\alpha\equiv\alpha(B)$, and the effective $g$-factor $g^{*}_{h}\equiv{}g^{*}_{h}(B)$ are all magnetic field dependent~\cite{RL_2021_Low}. This is the key difference between our hole `spin' and the previous hole spin models~\cite{PhysRevB.104.115425,PhysRevB.90.195421}. We note recently a similar magnetic field dependent hole spin model was also proposed in Ref.~\cite{PhysRevB.105.075308}. The two sets of combined subband dispersions in a transverse field (see Fig.~\ref{fig_fitting2}) are not well separated from each other, such that in the following we only consider the longitudinal field case. The detailed dispersion parameters in a longitudinal field are given in Tab.~\ref{tab1} in appendix~\ref{appendix_a}.

\begin{figure}
\includegraphics{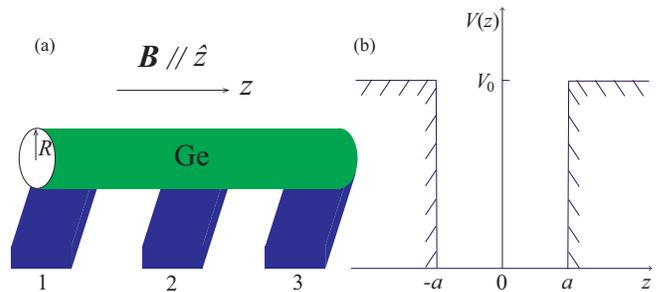}
\caption{\label{fig_model}(a) Possible realization of a gated nanowire hole quantum dot via placing a cylindrical Ge nanowire on top of a series of metallic gates. The nanowire has radius $R$ and the longitudinal direction is defined as the $z$ direction. A strong magnetic field is also applied in the $z$ direction. (b) A finite square well of size $a$ and height $V_{0}$ is used to model the longitudinal quantum dot confining potential. The quantum dot size $a$ is much larger than the nanowire radius $R$.}
\end{figure}

We now consider a semiconductor Ge nanowire quantum dot via introducing extra confining potential along the nanowire. For example, we can apply positive voltages to the metallic gates below the nanowire [see Fig.~\ref{fig_model}(a)]. When the quantum dot size $a$ [see Fig.~\ref{fig_model}(b)] in the longitudinal direction is much larger than the nanowire radius $R$ in the transverse direction, the Hamiltonian of the hole quantum dot is expected to be well represented by
\begin{equation}
H=\frac{\hbar^{2}k^{2}_{z}}{2m^{*}_{h}}+\alpha\sigma^{z}k_{z}+\frac{g^{*}_{h}\mu_{B}B}{2}\sigma^{x}+V(z),\label{eq_qdmodel}
\end{equation} 
where $k_{z}=-i\partial_{z}$ and a finite square well is used to model the quantum dot confining potential
\begin{equation}
V(z)=\left\{\begin{array}{cc}0, &  |z|<a, \\ V_{0}, & |z|\ge\,a,\end{array}\right.
\end{equation}
with $a$ and $V_{0}$ characterizing the width and the height of the well, respectively. We note that the nanowire hole quantum dot is always in the strong `spin'-orbit coupling regime, i.e., $m^{*}_{h}\alpha^{2}/\hbar^{2}>g^{*}_{h}\mu_{B}B/2$ is always satisfied for parameters given in Tabs.~\ref{tab1}. Also, in our following calculations, unless otherwise indicated, we have set $a=40$ nm and $V_{0}=5$ meV.

\section{\label{Sec_III}Electric-dipole `spin' resonance}

In this section, we study the electrical manipulation of a quantum dot `spin'-orbit qubit. Our first step is to obtain the lowest two energy levels in the nanowire quantum dot, to which the information of a hole `spin'-orbit qubit is encoded. We note that, even in the presence of the strong `spin'-orbit coupling, the 1D quantum dot Hamiltonian (\ref{eq_qdmodel}) is exactly solvable~\cite{lirui2018the}. A key step in the solution is to explore the $Z_{2}$ symmetry of the model, i.e., $(\sigma^{x}{\mathcal P})H(\sigma^{x}{\mathcal P})=H$, where $\mathcal{P}$ is the parity operator. It follows that we can classify the eigenstates $\Psi(z)=[\Psi_{1}(z),\Psi_{2}(z)]^{\rm T}$ in the quantum dot using this $Z_{2}$ symmetry. We find the ground state $\Psi_{\rm o}(z)$ always has symmetry $\sigma^{x}\mathcal{P}=-1$, while the first excited state $\Psi_{\rm e}(z)$ has symmetry $\sigma^{x}\mathcal{P}=1$. Hence, the Hilbert subspace of our hole `spin'-orbit qubit is spanned by $|\Psi_{\rm o}\rangle$ and $|\Psi_{\rm e}\rangle$. The detailed functional form of these two states can be found in Ref.~\cite{lirui2018the}.

We now consider the electric-dipole `spin' resonance where the hole `spin'-orbit qubit is manipulable by an external oscillating electric field. The total driving Hamiltonian of the hole quantum dot reads
\begin{equation}
H_{\rm t}=\frac{\hbar^{2}k^{2}_{z}}{2m^{*}_{h}}+\alpha\sigma^{z}k_{z}+\frac{g^{*}_{h}\mu_{B}B}{2}\sigma^{x}+V(z)+ezE\cos(2\pi\nu{t}),\label{eq_qddriving}
\end{equation}
where $E$ and $\nu$ are the amplitude and the frequency of the driving electric field, respectively. In the following, we have considered Eq.~(\ref{eq_qddriving}) for both the lowest and second lowest hole subband dispersions. The magnetic field dependences of parameters $m^{*}_{h}$, $\alpha$, and $g^{*}_{h}$ are given in Tab.~\ref{tab1} in appendix~\ref{appendix_a}. When the hole occupies the lowest subband dispersion, e.g., the quantum dot contains only one hole, the `spin'-orbit coupling $\alpha$ increases with the magnetic field $B$. When the hole occupies the second lowest subband dispersion, e.g., the quantum dot contains a few holes, the `spin'-orbit coupling $\alpha$ decreases with the magnetic field $B$. The opposite magnetic field dependences of the `spin'-orbit coupling of these two subband dispersions will lead to quite distinct results in the following.

Restricted to the Hilbert subspace spanned by the basis states $|\Psi_{\rm o}\rangle$ and $|\Psi_{\rm e}\rangle$ of the `spin'-orbit qubit, the Hamiltonian (\ref{eq_qddriving}) can be reduced to the Rabi oscillation Hamiltonian~\cite{scully1999quantum}
\begin{equation}
H_{\rm t}=\frac{1}{2}E_{\rm qu}\tau^{z}+h\Omega_{\rm R}\tau^{x}\cos(2\pi\nu{}t),\label{eq_EDSR}
\end{equation} 
where $E_{\rm qu}$ is the level splitting of the `spin'-orbit qubit, $\Omega_{\rm R}=|eE\langle\Psi_{\rm o}|z|\Psi_{\rm e}\rangle|/h$ is the Rabi frequency, with $h$ being the Planck constant, and $\tau^{z}=|\Psi_{\rm e}\rangle\langle\Psi_{\rm e}|-|\Psi_{\rm o}\rangle\langle\Psi_{\rm o}|$ and $\tau^{x}=|\Psi_{\rm e}\rangle\langle\Psi_{\rm o}|+|\Psi_{\rm o}\rangle\langle\Psi_{\rm e}|$ are the Pauli matrices. When the frequency of the driving electric field matches the qubit level splitting $E_{\rm qu}=h\nu$, we are able to achieve a coherent `spin'-orbit qubit manipulation. In terms of the qubit density matrix, Eq.~(\ref{eq_EDSR}) will give rise to periodic oscillation of the qubit inversion $W(t)=\rho_{\rm e,e}(t)-\rho_{\rm o,o}(t)$~\cite{scully1999quantum}, since we do not consider any external environmental noise.

\begin{figure}
\includegraphics{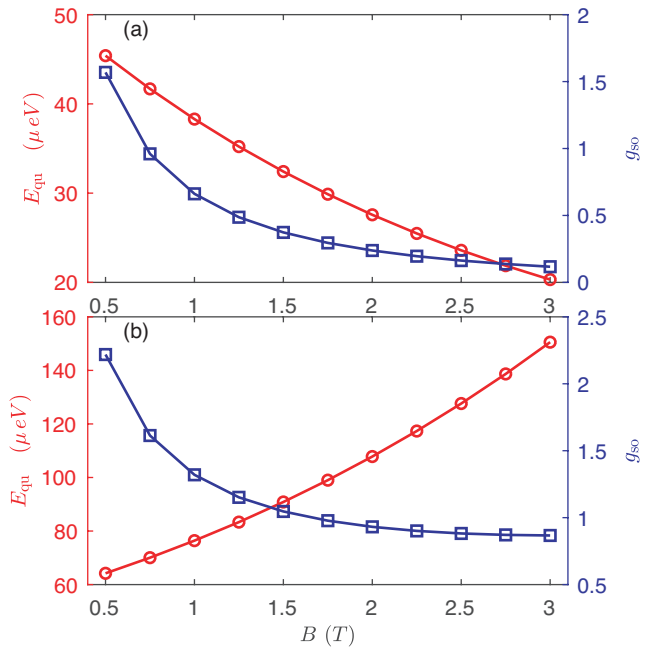}
\caption{\label{fig_levelsplit_L}Both the level splitting $E_{\rm qu}$ and the $g$-factor $g_{\rm so}$ of the hole `spin'-orbit qubit as function of the longitudinal magnetic field $B$.  (a) The results for the lowest subband dispersion. (b) The results for the second lowest subband dispersion.}
\end{figure}

\begin{figure}
\includegraphics{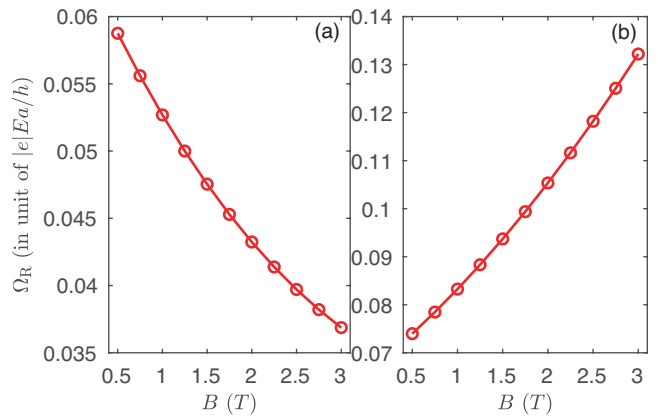}
\caption{\label{fig_rabif_L}The Rabi frequency $\Omega_{R}$ as a function of the longitudinal magnetic field $B$. For a moderate ac electric field amplitude $E=10^3$ V/m, the Rabi frequency unit is $|e|Ea/h=9.67$ GHz. (a) The result for the lowest subband dispersion. (b) The result for the second lowest subband dispersion.}
\end{figure}

We show the level splitting of the hole `spin'-orbit qubit as a function of the applied longitudinal field in Fig.~\ref{fig_levelsplit_L}. In order to  reveal explicitly the impact of the `spin'-orbit coupling on the qubit level splitting, we define a `spin'-orbit coupling modulated $g$-factor, i.e, the $g$-factor of the `spin'-orbit qubit
\begin{equation}
g_{\rm so}=E_{\rm qu}/(\mu_{B}B). 
\end{equation} 
Obviously, without the `spin'-orbit coupling in Hamiltonian (\ref{eq_qdmodel}), $g_{\rm so}$ would equal to $g^{*}_{h}$. Due to the presence of the strong `spin'-orbit coupling, $g_{\rm so}$ is much less than $g^{*}_{h}$ (see both Fig.~\ref{fig_levelsplit_L} and Tab.~\ref{tab1}). Note that a recent experiment also reported a non-vanishing longitudinal $g$-factor $g_{\rm so}=1.06$~\cite{Froning:2021aa} in the weak magnetic field region ($B<0.4$ T). The level splitting $E_{\rm qu}$ of the lowest subband dispersion decreases with the field $B$ [see Fig.~\ref{fig_levelsplit_L}(a)], while the level splitting $E_{\rm qu}$ of the second lowest subband increases with the field $B$ [see Fig.~\ref{fig_levelsplit_L}(b)]. For both these two subband dispersions, the `spin'-orbit coupling modulated $g$-factors $g_{\rm so}$ decrease with the field $B$. These results are understandable if we borrow the analytical result $g_{\rm so}=g^{*}_{h}{\rm exp}(-a^{2}/z^{2}_{\rm so})$~\cite{trif2008spin,lirui2013controlling} when $V_{z}$ is a harmonic potential. When the magnetic field $B$ increases, the spin-orbit length $z_{\rm so}=\hbar^{2}/(m^{*}_{h}\alpha)$ has only a very small magnetic field dependence, while the hole $g$-factor $g^{*}_{h}$ decreases significantly (see Tab.~\ref{tab1}), such that it is reasonable $g_{\rm so}$ decreases with the increase of the magnetic field.


The Rabi frequency is an important physical quantity characterizing the manipulation time of the hole `spin'-orbit qubit. We show the Rabi frequency as a function of the longitudinal field in Fig.~\ref{fig_rabif_L}. Figures~\ref{fig_rabif_L}(a) and (b) show the results of the lowest and second lowest hole subband dispersions, respectively. The Rabi frequency of the lowest subband dispersion decreases with the magnetic field $B$, while it increases with the magnetic field $B$ for the second lowest subband dispersion. Note that the Rabi frequency of the lowest subband dispersion can be as large as $0.059|e|Ea/h$ ($\sim568$ MHz) given at $B=0.5$ T. The frequencies of the driving electric field in experiments are usually restricted to the interval $1-20$ GHz, such that the maximum detectable Rabi frequency of the second lowest subband dispersion is about $0.083|e|Ea/h$ ($\sim800$ MHz) given at $B=1$ T.

\section{\label{Sec_IV}The effects of varying the well height}
\begin{figure}
\includegraphics{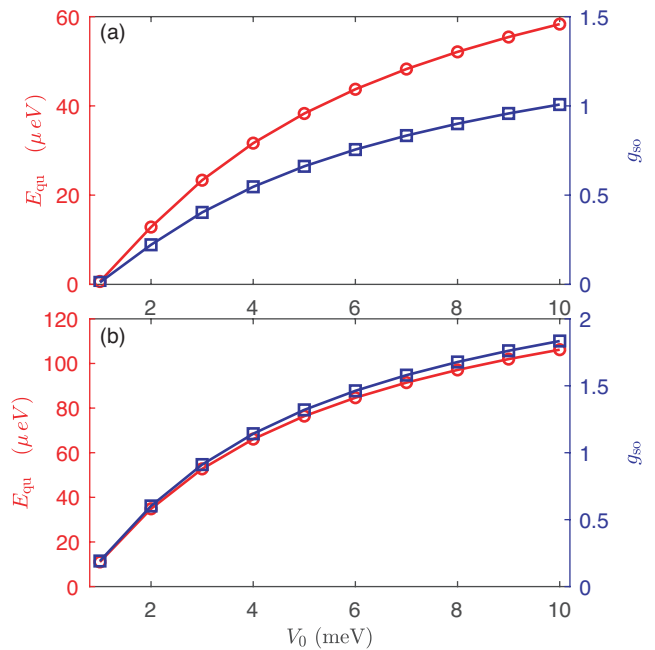}
\caption{\label{fig_varyV0I}Both the level splitting $E_{\rm qu}$ and the $g$-factor $g_{\rm so}$ of the hole `spin'-orbit qubit as a function of the  potential height $V_{0}$. The magnetic field is fixed longitudinally at $B=1$ T. (a) The result for the lowest subband dispersion. (b) The result for second lowest subband dispersion.}
\end{figure}

\begin{figure}
\includegraphics{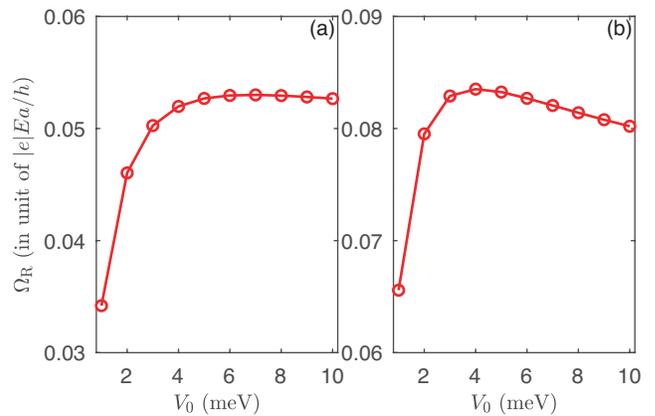}
\caption{\label{fig_varyV0II}The Rabi frequency $\Omega_{\rm R}$ as a function of the potential height $V_{0}$. The magnetic field is fixed longitudinally at $B=1$ T. (a) The result for the lowest subband dispersion. (b) The result for second lowest subband dispersion.}
\end{figure}

In the previous considerations, we have set the height and the size of the square well to $V_{0}=5$ meV and $a=40$ nm, respectively. Here we study the effects of varying the well height $V_{0}$ (keep the well size $a$ unchanging during this process). This is highly relevant to the experimental circumstance. When a gated nanowire quantum dot is fabricated experimentally, the distances between the local gates below the nanowire are usually also fixed, such that it may be difficult to tune the quantum dot size $a$. However, it is easy to tune the height of the confining potential $V_{0}$ via varying the corresponding gate voltage, e.g., the voltage on gate `2' in Fig.~\ref{fig_model}(a).

As a concrete example, in our following calculations, we fix the longitudinal magnetic field at $B=1$ T. When the hole occupies the lowest hole subband dispersion, both the level splitting $E_{\rm qu}$ and the $g$-factor $g_{\rm so}$ of the `spin'-orbit qubit as a function of the potential height $V_{0}$ are shown in Fig.~\ref{fig_varyV0I}(a). When the hole occupies the second lowest hole subband dispersion, the corresponding results are shown in Fig.~\ref{fig_varyV0I}(b). For these two subband dispersions, when we reduce the height of the confining potential $V_{0}$, the level splitting of the hole `spin'-orbit qubit decreases. This response is very reasonable, because when we reduce $V_{0}$, we effectively increase the characteristic length $a_{\rm ef}$ of the nanowire quantum dot. We can imagine that for smaller height $V_{0}$, the eigenstate will penetrate more to the potential barrier given in the coordinate region $|z|>a$, such that the effective quantum dot characteristic length is increased ($a_{\rm ef}>a$). It follows that the relative `spin'-orbit coupling characterized by $a_{\rm ef}/z_{\rm so}$ in the nanowire quantum dot is also increased.

We show the Rabi frequency $\Omega_{R}$ as a function of the potential height $V_{0}$ for the lowest and second lowest subband dispersions in Figs.~\ref{fig_varyV0II}(a) and \ref{fig_varyV0II}(b), respectively. In both cases, there exists a maximum of the Rabi frequency when we reduce the potential height $V_{0}$. This phenomenon has been observed previously~\cite{lirui2018the}. We can enhance the `spin'-orbit coupling effect in the nanowire quantum dot via reducing the confining potential height. After reaching the maximum, if we still reduce $V_{0}$, the Rabi frequency decreases sharply. Anyway, it is possible to tune both the `spin'-orbit coupling modulated $g$-factor $g_{\rm so}$ and the Rabi frequency $\Omega_{\rm R}$ to some extent through the quantum dot gate voltage.

\section{\label{Sec_V}Discussion and summary}

In this paper, we have only elaborately considered the longitudinal magnetic field case. Here, let us give a brief discussion on the potential results when the external magnetic field is transverse. The hole subband dispersions in a transverse field are shown in Fig.~\ref{fig_fitting2} in appendix~\ref{appendix_a}. There still exist two sets of combined subband dispersions, i.e., the solid line and the dashed-dot line in Fig.~\ref{fig_fitting2}, each of which can still be modeled by Eq.~(\ref{eq_dispersion}). However, these two combined subband dispersions are not well separated from each other. Nevertheless, for each combined subband dispersion in Fig.~\ref{fig_fitting2}, we still can have similar results as those given in Figs.~\ref{fig_levelsplit_L} and \ref{fig_rabif_L}.

Note that the hole `spin' studied in this paper is essentially different from the real hole spin. The real hole spin splitting in a magnetic field is the splitting between the solid line and the dashed dot line in Figs.~\ref{fig_fitting1} and \ref{fig_fitting2} in appendix~\ref{appendix_a}. The longitudinal $g$-factor of the real hole spin at $k_{z}R=0$ is very small $g_{l}\approx0.14$~\cite{PhysRevB.84.195314,RL_2021_Low,PhysRevB.93.121408,PhysRevB.87.161305}, while the longitudinal $g$-factor of the hole `spin' can have significant values (see Tab.~\ref{tab1})~\cite{Li_2022_Searching}. Thus, in terms of the hole `spin', we are able to produce both a non-vanishing `spin'-orbit coupling modulated longitudinal $g$-factor and a large electrical Rabi frequency which may be relevant to the recent experimental observation~\cite{Froning:2021aa}, where a non-vanishing longitudinal $g$-factor $g^{l}_{\rm so}=1.06$ is observed in the low magnetic field region ($B<0.4$ T). Note that in our considered magnetic field interval, the lowest two energy levels in the quantum dot indeed represent a hole `spin'-orbit qubit, instead of a real hole spin. Also, the magnetic fields considered in this paper are experimentally achievable~\cite{Scappucci:2021vk,Hendrickx:2020ab}.

In summary, in this paper, we have studied in details the magnetic field effects in the electric-dipole `spin' resonance in a Ge nanowire hole quantum dot. The effective hole mass $m^{*}_{h}$, the `spin'-orbit coupling $\alpha$, and the effective $g$-factor $g^{*}_{h}$ are all magnetic field dependent, such that both the level splitting and the Rabi frequency of the `spin'-orbit qubit are shown to have non-linear and non-trivial dependences on the magnetic field strength. Also, the results of the lowest hole subband are totally different from that of the second lowest hole subband. The qubit level splitting $E_{\rm qu}$ of the lowest subband decreases with the magnetic field, while it increases with the magnetic field for the second lowest subband. The $g$-factor of the `spin'-orbit qubit $g_{\rm so}$ decreases with the magnetic field for both subband dispersions. We finally show the tunabilities of both the $g$-factor $g_{\rm so}$ and the Rabi frequency $\Omega_{\rm R}$ of the `spin'-orbit qubit by varying the quantum dot confining potential. Our paper demonstrates that the hole `spin' manipulation frequency of several hundreds of MHz is achievable in the Ge nanowire quantum dot.

\section*{Acknowledgements}
This work was supported by the National Natural Science Foundation of China Grant No.~11404020, the Project from the Department of Education of Hebei Province Grant No. QN2019057, and the Starting up Foundation from Yanshan University Grant No. BL18043.

\appendix
\section{\label{appendix_a}The magnetic field dependences of the quantum dot parameters}
\begin{table*}
	\centering
	\caption{\label{tab1}The magnetic field dependences of the effective hole mass $m^{*}_{h}$, the `spin'-orbit coupling $\alpha$, and the effective $g$-factor $g^{*}_{h}$. The magnetic field ${\bf B}=(0,0,B)$ is applied longitudinally and the nanowire radius is $R=10$ nm.}
	\begin{ruledtabular}
		\begin{tabular}{c|ccc|ccc}
		         ~&\multicolumn{3}{c|}{The lowest subband}&\multicolumn{3}{c}{The second lowest subband}\\\hline
			$B$~(T)&$m^{*}_{h}/m_{e}$\footnote{$m_{e}$ is the free electron mass}&$\alpha$~(eV \AA)&$g^{*}_{h}$&$m^{*}_{h}/m_{e}$&$\alpha$~(eV \AA)&$g^{*}_{h}$\\
			$0.5$&$0.07345$&$0.55354$&$19.27596$&$0.07425$&$0.53257$&$22.28171$\\
			$0.75$&$0.07326$&$0.55867$&$12.43149$&$0.07446$&$0.52721$&$15.43724$\\
			$1$&$0.07307$&$0.56375$&$9.03379$&$0.07467$&$0.52180$&$12.03955$\\
			$1.25$&$0.07288$&$0.56879$&$7.01482$&$0.07488$&$0.51634$&$10.02057$\\
			$1.5$&$0.07270$&$0.57379$&$5.68521$&$0.07510$&$0.51083$&$8.69096$\\
			$1.75$&$0.07252$&$0.57874$&$4.74953$&$0.07533$&$0.50527$&$7.75526$\\
			$2$&$0.07234$&$0.58365$&$4.06005$&$0.07555$&$0.49966$&$7.06577$\\
			$2.25$&$0.07217$&$0.58852$&$3.53473$&$0.07579$&$0.49400$&$6.54041$\\
			$2.5$&$0.07199$&$0.59334$&$3.12431$&$0.07602$&$0.48828$&$6.12995$\\
			$2.75$&$0.07182$&$0.59812$&$2.79746$&$0.07627$&$0.48250$&$5.80304$\\
			$3$&$0.07166$&$0.60286$&$2.53329$&$0.07651$&$0.47667$&$5.53880$\\
		\end{tabular}
	\end{ruledtabular}
\end{table*}

\begin{figure}
\includegraphics{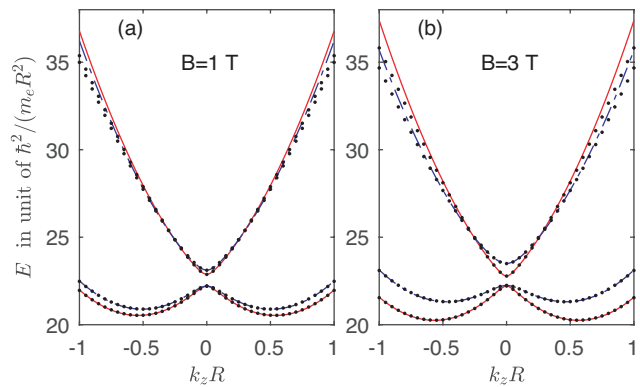}
\caption{\label{fig_fitting1}Fitting to the data of the hole subband dispersions in a longitudinal field by using Eq.~(\ref{eq_dispersion}). (a) The result for magnetic field $B=1$ T.  (b) The result for magnetic field $B=3$ T. The fitting parameters $m^{*}_{h}$, $\alpha$, and $g^{*}_{h}$ can be found in Tab.~\ref{tab1}.}
\end{figure}
\begin{figure}
\includegraphics{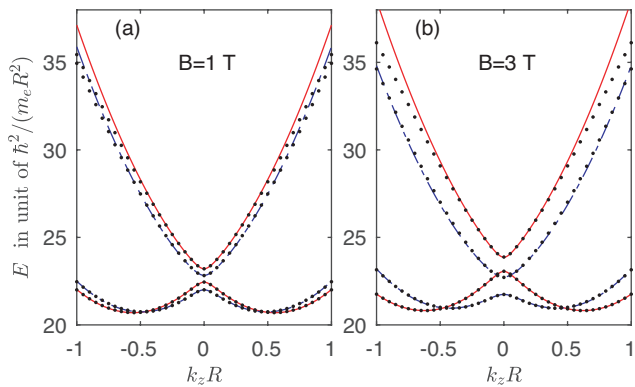}
\caption{\label{fig_fitting2}The hole subband dispersions in a transverse field. (a) The result for magnetic field $B=1$ T. (b)The result for magnetic field $B=3$ T.}
\end{figure}

In this appendix, we give the detailed parameters of the Ge nanowire quantum dot when the external magnetic field is varying. When the external magnetic field is applied longitudinally (parallel to the nanowire), the magnetic field dependences of the quantum dot parameters are given in Tab.~\ref{tab1}. The values of these parameters are determined by fitting Eq.~(\ref{eq_dispersion}) to the data of the subband dispersions, which are obtained using quasi-degenerate perturbation theory~\cite{Li_2022_Searching}. The fitting results are shown in Fig.~\ref{fig_fitting1}. There are two sets of combined subband dispersions, i.e., the solid line and the dashed-dot line in Fig.~\ref{fig_fitting1}, which are labeled as the lowest and second lowest subbands, respectively. Also, with the increase of the longitudinal field, these two subbands become more separated from each other  [see Figs.~\ref{fig_fitting1}(a) and (b)]. When the external magnetic field is applied transversely (perpendicular to the nanowire), the hole subband dispersions are given in Fig.~\ref{fig_fitting2}. There still exist two sets of combined subband dispersions, each of which can still be modeled by Eq.~(\ref{eq_dispersion}). However, these two subbands are not well separated from each other, such that in this paper we do not give the detailed calculations for the transverse field case.

\bibliographystyle{iopart-num}
\bibliography{Ref_Hole_spin}

\end{document}